\documentclass[10pt,conference]{IEEEtran}

\usepackage{amsfonts,amssymb,amsmath,graphics,epsfig}

\newtheorem{thm}{Theorem}
\newtheorem{lemma}{Lemma}

\begin{document}

\def\endproof{\hfill$\Box$\vspace{0.4cm}}

\def\ve{\varepsilon}
\def\cG{{\cal G}}
\def\us{\underline{s}}
\def\ux{\underline{x}}
\def\uy{\underline{y}}
\def\uw{\underline{w}}
\def\expect{{\mathbb E}}
\def\<{\langle}
\def\>{\rangle}
\def\ed{\stackrel{{\rm d}}{=}}

\def\reals{{\mathbb R}}
\def\sig{{\mathbb S}}
\def\de{{\rm d}}
\def\bit{{\rm P}_{\rm b}}
\def\E{{\mathbb E}}
\def\Ez{{\sf E}_z}
\def\la{\overline{l}}
\def\prob{{\mathbb P}}
\def\Var{{\rm Var}}

\def\hs{\overline{h}}
\def\hi{\underline{h}}

\def\gexit{g}
\def\gbp{g_{\mbox{\tiny BP}}}
\def\lbp{\lambda_{\mbox{\tiny BP}}}
\def\qbp{q_{\mbox{\tiny BP}}}
\def\ex{\widehat{x}}

\def\hrs{h_{\mbox{\tiny RS}}}

\date{\today}

\title{Analysis of Belief Propagation for Non-Linear Problems:\\
The Example of CDMA (or: How to Prove Tanaka's Formula)}

\author{\authorblockN{Andrea Montanari}
\authorblockA{Laboratoire de Physique Th\'{e}orique\\ de l'Ecole Normale
Sup\'{e}rieure,\\ CNRS-UMR 8549\\
Email: {\tt montanar@lpt.ens.fr}}
\and
\authorblockN{David Tse}
\authorblockA{Department of Electrical Engineering\\ and Computer Science,\\
University of California, Berkeley\\
Email: {\tt dtse@eecs.berkeley.edu}} 
 }
\maketitle

\begin{abstract}
We consider the CDMA (code-division multiple-access) multi-user detection
problem for binary signals and additive white gaussian noise.
We propose a spreading sequences scheme based on random sparse signatures,
and a detection algorithm based on belief propagation (BP) with linear
time complexity. In the new scheme, each user conveys its power
onto a {\em finite} number of chips $\la$, in the large system limit. 

We analyze the performances of BP detection and prove that they
coincide with the ones of optimal (symbol MAP) detection in the
$\la\to\infty$ limit. In the same limit, we prove that the information
capacity of the system converges to Tanaka's formula for random
`dense' signatures, thus providing the {\em first} rigorous justification 
of this formula. Apart from being computationally convenient, the 
new scheme allows for optimization in close analogy with
irregular low density parity check code ensembles.
\end{abstract}

%
%
\section{Introduction}

\subsection{Motivation}

The crucial new characteristics of modern (iterative)
coding systems \cite{RichardsonUrbankeBook}
are: $(i)$ Probabilistic construction based on
sparse random graphs; $(ii)$ Iterative (belief propagation, BP)
decoding; $(iii)$ Focus onto the large system limit. 
Despite their generality, the impact of these principles outside 
the area of linear error correcting codes has been limited.
It is therefore extremely interesting to extend their scope to 
other communications and information theory 
problems\footnote{An earlier example that support this view is
the use of low density codes with non-linear checks for lossy 
data compression in \cite{Stefano}.}.

The tools developed for the analysis of iterative coding systems
must be considerably strenghtened in order to cope with such generalizations. 
Consider for instance the  question
of whether BP decoding is asymptotically optimal (in the large
system limit), i.e. if it implements symbol MAP decoding. 
For LDPC codes, density evolution (DE) allows to show
that this is the case if the  noise level is smaller than a threshold, 
below which the asymptotic BP bit error rate $\bit^{\rm BP}$ vanishes. 
When  $\bit^{\rm BP}>0$ (as we expect in a general setting), 
one cannot say much about 
MAP performances, and their relation to BP (apart from the obvious 
sub-optimality of BP). 

Recently, some definite progress was made on these problems in the context
of LDPC codes \cite{Life,BECArea,GeneralArea}. 
The basic new ingredient  is a `general area theorem'
that yields the rate of change of the mutual information across 
the system, under a change in the channel parameter.
Earlier examples of such a relation were found by 
Ashikhmin, Kramer , and ten Brink \cite{Ashikhmin} (for the erasure channel), 
and Guo, Shamai and Verd\`u \cite{GSV,GSVP} (for the gaussian and Poisson
channels).
The approach based on the area theorem seems rather general. 
In order to illustrate it, and further explore its capabilities,
we consider here a new application: multi-user detection \cite{VerduBook}.
%
%
\subsection{Multi-user detection with binary inputs}

In a simple multi-user detection scenario, each of $K$ users transmits a
symbol $x_i\in\reals$ to a common receiver, after encoding it 
using a signature $\us_i\in\reals^N$. The received signal is
\begin{eqnarray}
\uy = \sum_i x_i\, \us_i +\uw\, .
\end{eqnarray}
where the noise $\uw$ is a vector of $N$ i.i.d. gaussian variables
of mean $0$ and variance $\sigma^2$. The input symbols $x_i$ are
also modeled as i.i.d.'s.
Writing $\sig$ for the $N\times K$ matrix with columns
$\us_1, \dots$, $\us_K$, and $x =(x_1,\dots,x_K)^{\rm T}$ for the input,
the above equation can also be written
$\uy=\sig\, \ux+\uw$.
Of great interest is the
large system limit $N,K\to\infty$ with $K/N=\alpha$ fixed.

How reliably can the input $\ux$
 be reconstructed given $\uy$ and the signature
matrix $\sig$? 
In order to answer this question, the signatures $\us_i$ are usually taken 
to be i.i.d. random vectors. The standard choice is to set 
$\us_i = \frac{1}{\sqrt{N}}  (s_{i1},\dots,s_{iN})^{\rm T}$ where the 
$s_{ia}$ are i.i.d. with zero mean and unit variance (we will call these
`dense signatures').
Tse and Hanly \cite{TseHanly}, and Verd\`u and Shamai \cite{VerduShamai} 
considered the case in which the input symbols $x_i$ are gaussian 
random variables. Using random matrix theory, they were able 
to compute the minimum mean square error, and the information capacity
of the system. In \cite{OursGaussian}, we considered a multi-user 
detection algorithm based on BP, and proved it to be optimal 
(i.e. to implement minimum mean square error detection) with high probability
in the large system limit.

The case of binary input symbols $x_i\in\{+ 1,-1\}$ uniformly at random,
is of obvious interest for practical applications, and out of reach
of classical methods (such as random matrix theory).
Tanaka \cite{Tanaka} 
used the replica method from statistical physics in order to determine
the asymptotic information capacity. More precisely, let us define
per-user conditional entropy $h\equiv\lim_{K\to\infty}K^{-1}\E H(X|Y)$,
where the expectation is taken with respect to the random signatures
and throughout the paper we measure entropies in nats
(obviously $I(X;Y) = K\log 2-H(X|Y)$).
He obtained $h=\hrs(\sigma^2,\alpha)\equiv \sup_q\hrs
(q;\sigma^2,\alpha)$, where
\begin{eqnarray}
\hrs(q;\sigma^2,\alpha) 
&=& \Ez\log 2\cosh(\lambda(q)+\sqrt{\lambda(q)}\, z)-\label{hrs}\\
&&-\frac{1}{2}\lambda (1+q)
-\frac{1}{2\alpha}\log\left(1+\frac{\alpha}{\sigma^2}(1-q)\right)\, ,
\nonumber
\end{eqnarray}
$\lambda(q) = [\sigma^2+\alpha(1-q)]^{-1}$, and $\Ez$ denotes throughut the 
paper expectation with respect to the standard normal variable $z$. 
It is easy to show that the value of $q$ maximizing $\hrs(q;\sigma^2,\alpha)$ 
must satisfy the 
stationarity condition 
\begin{eqnarray}
q = \Ez \tanh^2(\lambda(q)+\sqrt{\lambda(q)}\, z)\, .\label{eq:SaddlePoint}
\end{eqnarray}
Unhappily, the replica method is non-rigorous.
In this paper we will prove Tanaka's formula for
$\alpha\le\alpha_{\rm s}\approx 1.49$ (a precise definition of 
$\alpha_{\rm s}$ is provided in the next Section).
For earlier applications of BP to multi-user detection with binary 
signals, we refer, for instance to \cite{Kabashima,TanakaOkada,NeirottiSaad}.
We will prove that, in the same regime $\alpha<\alpha_{\rm s}$, 
optimal (symbol MAP) detection can be implemented using BP. 

In order to prove these results, we will introduce a new 
`sparse signature' scheme, see Section \ref{SchemeSection}, and
view standard dense signatures as a limiting case.
The identity between the two limiting procedures will be the object of
a separate publication.
The new scheme (which is reminiscent of LT codes \cite{LT}) 
is on the other hand interesting {\em per se}. It allows to implement 
BP in a very natural way with complexity linear in $N$. Furthermore, 
it opens the way to optimization of the degree sequence thus improving the
performances over dense signatures. We refer to Section \ref{NumericalSection}
for numerical indications in this direction.
%
%
\section{The sparse signature scheme, and main results}
\label{SchemeSection}

\subsection{Sparse signatures and belief propagation}

As already mentioned, in order to prove Tanaka's formula we
shall introduce a new signature scheme. This is caracterized by a distribution 
$\{\Omega_l: l\ge 0\}$ over the non negative integers (to avoid 
pathological behaviors, we assume it to have bounded support). 
We also let $\la>0$ be its mean and define $\omega_l \equiv l\Omega_l/\la$
for $l\ge 0$.
The user $i$ constructs her signature $s_i$ independently from the other users
as follows. She chooses an integer $l$ from the distribution 
$\Omega_l$, and a subset $\partial i$ of $\{1,\dots,N\}$ of
size $|\partial i|=l$ uniformly at random among the $\binom{N}{l}$ such 
subsets. Her signature is $s_i = 
\frac{1}{\sqrt{\la}}(s_{i1},\dots,s_{iN})^{\rm T}$ where 
$s_{ia}\in\{+1,-1\}$
uniformly at random if $a\in\partial i$, and $s_{ia}=0$ otherwise.

\begin{figure}
\center{\includegraphics[width=5cm]{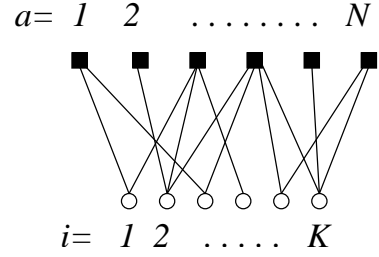}}
\caption{Factor graph representation of the sparse signature scheme:
circles represent users (variable nodes) and squares chips (function nodes).}
\label{fig:factor}
\end{figure}
Notice that the normalization ensures that the average power 
employed by each user is equal to $1$ as for the dense signature 
scheme. However this power is conveyed onto a finite number of chips.
Viceversa, each chip $a\in\{1,\dots,N\}$ receives power from a finite
number of users, to be denoted as $\partial a$ (this is the set of
$i\in \{1,\dots,K\}$ such that $a\in\partial i$).
The conditional distribution of the input symbols, given the received
signal $y$ take the form
\begin{eqnarray}
\mu^{y,{\mathbb S}}(x) &= 
&\frac{1}{Z}\, \prod_{a=1}^N\psi_{y_a}(x_{\partial a})\, ,\\
\psi_{y_a}(x_{\partial a}) &= &\, 
\exp\left\{-\frac{1}{2\sigma^2}\left(y_a-\sum_{i\in\partial a}
\frac{s_{ia}}{\sqrt{\la}}x_i\right)^2\right\}\, .
\end{eqnarray}
Such distribution is conveniently represented through the associated 
factor graph, cf. Fig.~\ref{fig:factor}.
This includes $K$ variable nodes (one for each user $i$),
$N$ function nodes (one for each chip $a$) and an edge joining
user $i$ and chip $a$ whenever $i\in\partial a$.

If signatures are chosen  according to the proposed scheme, 
the resulting factor graph is a sparse random graph. 
The degree distribution is $\Omega_l$ on the
variable node (user) side, and converges to a Poisson distribution with mean 
$\la \alpha$ on the function node (chip) side.

BP is introduced in the standard way:
we limit ourselves to writing down the update equations
in terms of log-likelihoods\footnote{More precisely, we use here {\em one
half} of log-likelihoods.}. Two types of messages are updated:
variable to function node, $v_{i\to a}$, and function to variable node,
$u_{a\to i}$. The update equations read
\begin{eqnarray}
v_{i\to a}^{t}  &=&\sum_{b\in\partial i\backslash a}
u_{b\to i}^{t-1}\, ,\label{eq:BP1}\\
u_{a\to i}^t &=& f(v_{j\to a}, s_{ja}, j\in\partial a\backslash i;\, s_{ia}
;\,y_a)\, ,\label{eq:BP2}
\end{eqnarray}
where the index $t$ denotes the iteration number and 
\begin{align}
&f(v_1,s_1,\dots,v_k,s_k;\,s_0;\,y) \equiv\frac{1}{2}\log
\frac{W_+}{W_-}\, ,\\
&W_{\xi_0}\equiv \sum_{\xi_1\dots\xi_k=\pm 1}
e^{-\frac{1}{2\sigma^2}\left(y-\sum_{i=0}^k
\frac{s_{i}}{\sqrt{\la}}\xi_i\right)^2}
\prod_{i=1}^k e^{v_i\xi_i}\, .
\end{align}
We furthermore adopt the initial condition $u^{0}_{a\to i} = v^{0}_{i\to
  a}=0$. After a fixed number of iterations, all the messages incoming 
at variable node $i$ are combined to compute the 
decision $x^{\mbox{\tiny BP},t}_i \equiv 
\text{sign}\{\sum_{a\in \partial i}u_{a\to i}^t\}$.
%
%
\subsection{Main results}
\label{ResultsSection}

In order to state and prove our main results more easily, it is 
convenient to focus onto `Poisson' signature schemes. By this we 
mean that  $\{\Omega_l, l\ge 0\}$ is a Poisson distribution
of mean $\la$. We shall come back to
the general case in Sections \ref{RemarksSection} and \ref{NumericalSection}.
Within this setting, we consider the expected conditional entropy per 
user $\E\, H(X|Y)/K$ (the expectation being taken with respect to
the random signatures). Since we do not know {\em a priori} whether the 
large system limit exists, we define
$\hs(\sigma^2,\alpha,\la) \equiv \limsup_{N\to\infty}\,\E\,H(X|Y)/K$, and 
$\hi(\sigma^2,\alpha,\la) \equiv \liminf_{N\to\infty}\,\E\,H(X|Y)/K$.
In both cases, the limit is taken keeping the ratio
$K/N=\alpha$ fixed.

If we let $\la\to N$ and then $N\to\infty$, we would recover
the standard dense signature scheme (strictly speaking this
corresponds to $\Omega_l$  concentrated on $l=N$). 
Here we shall invert the order
of the two limits and let $N\to \infty$ and then $\la\to\infty$
{\em afterwards}. Our first result shows that, if the limit is taken in this 
way,  Tanaka formula is correct. For our proof technique to work
$\alpha$ must be smaller than the `spinodal value' $\alpha_{\rm s}$.
This is the largest number such that, for any $\alpha<\alpha_{\rm s}$
the solution to Eq.~(\ref{eq:SaddlePoint}) is unique for all 
$\sigma^2\in[0,\infty)$, and is a differentiable 
function of $\sigma^2$. By solving Eq.~(\ref{eq:SaddlePoint}) 
numerically, we get $\alpha_{\rm s}\approx 1.49$.
\begin{thm}\label{thm:Entropy}
If $\alpha<\alpha_{\rm s}$, then the  per-user conditional entropy
converges to Tanaka's formula in the dense signature limit
\begin{eqnarray}
\lim_{l\to\infty} \hs(\sigma^2,\alpha,\la) = \lim_{l\to\infty} 
\hi(\sigma^2,\alpha,\la) = \hrs(\sigma^2,\alpha)\, . 
\end{eqnarray}
\end{thm}
The hypothesis of Poisson signatures is presently used only in 
the proof of Lemma \ref{lemma:ZeroNoise}. 
It shouldn't however be difficult to extend this result
to more general sequences of degree distributions $\Omega_l$. 

A key step in the proof of the above result consists in analizing
the BP-based detection algorithm defined by Eqs.~(\ref{eq:BP1}),
(\ref{eq:BP2}). Our second result shows that, in the 
small $\alpha$ regime this algorithm is indeed optimal (the proof of
this result is  deferred to a longer paper).
\begin{thm}\label{thm:BP}
Let $\overline{\bit}(\la,N)$ be the expected bit error rate under 
symbol MAP detection, and $\overline{\bit^{\mbox{\tiny BP}}}(\la,N;t)$
the same quantity for $t$ iterations BP detection. Define the
asymptotic BP error overhead as 
\begin{eqnarray}
\Delta(\la;t) = \limsup_{N\to\infty}
[\overline{\bit^{\mbox{\tiny BP}}}(\la,N;t)-\overline{\bit}(\la,N)]\, .
\end{eqnarray}
If $\alpha<\alpha_{\rm s}$, then BP is optimal in the dense signature limit,
namely $\lim_{t\to\infty}\lim_{\la\to\infty}\Delta(\la;t) = 0$.
\end{thm}
%
%
\section{A sketch of the proof}
\label{ProofSection}

%
%
\subsection{A few simple remarks}
\label{RemarksSection}

We start by collecting a few remarks whose proof is routine,
and therefore omitted apart from a few hints.
\vspace{0.1cm}

{\em \underline{All $+1$ input.}} 
For the sake of analysis 
(and for proving Theorem \ref{thm:Entropy}) we can assume that 
the input signal is $x = x_+ \equiv (+1,\dots,+1)^{\rm T}$.
In particular, if we write $\E_{y,\sig}^{+}$ for the joint 
expectation with respect to $y$ and $\sig$, conditional to 
$x = x_+$, then $\E_{\sig} H(X|Y) = 
-\E_{x,y,\sig} \log \prob(X|Y,\sig) =
-\E^+_{y,\sig}\log\prob(X=x_+|Y,\sig)$.
\vspace{0.1cm}

{\em \underline{Density evolution.}}(DE) Any finite neighborhood of a randomly 
chosen node in the factor graph associated to the sparse signature scheme,
converges in distribution to a tree with the degree distribution mentioned
above. As a consequence, the messages distribution can be analyzed 
through a standard DE approach. 

Define the sequence of random variables $\{v^t,u^t;\, t\ge 0\}$ as follows:
$v^0 = u^0= 0$, and
\begin{eqnarray}
v^{t+1}  \ed\sum_{b=1}^{l}
u_{b}^{t}\, ,\;\;\;\;
u^t \ed f(v^t_{1}, s_{1}, \dots,v^t_k,s_k;\, s_{0}
;y)\, ,
\end{eqnarray}
for $t\ge 0$. Here $\ed$ denotes identity in distribution;
$u^{t}_1,u^{t}_2,\dots$ (respectively, $v^{t}_1,v^{t}_2,\dots$)
are i.i.d. copies of $u^t$ (respectively, of $v^t$);
$l$ is an integer random variable with distribution $\omega_l$,
and $k$ is a Poisson random variable with mean $\la\alpha$;
finally $s_0,\dots,s_k$ are i.i.d.'s with $s_i\in\{+ 1,-1\}$
uniformly at random, $y= \frac{1}{\sqrt{\la}}\sum_{i=0}^ks_i+w$ with
$w$ a normal random variable with mean $0$ and variance $\sigma^2$.

Let $(ia)$ be a uniformly random edge in the factor graph
and $v_{i\to a}^t$, $u^t_{a\to i}$ the corresponding BP messages,
under the assumption that $x_+$ has been transmitted. 
Then $v_{i\to a}^t$ (respectively $u_{a\to i}^t$) converges in 
distribution to $v^t$ (respectively, to $u^t$) as $N\to\infty$.
\vspace{0.1cm}

{\em \underline{Symmetry condition.}}
 A random variable $X$ is `symmetric' if
$\E[f(-X)] =\E[\; e^{-2X}f(X)]$
for any function $f$ such that both expectation exist.
It is easy to show that the random variables $u^t$, $v^t$ defined above are
symmetric (this is analogous to what happens in LDPC codes).
\vspace{0.1cm}

{\em \underline{Area theorem.}} Following \cite{GSV}, the derivative,
with respect to the noise parameter, of the conditional entropy
is proportional to the expectation of the conditional variance
\begin{eqnarray}
\frac{\de H(X|Y)}{\de \sigma^2} = \frac{1}{2\sigma^4}
\E_y\left\{\Var(\sig X|Y)\right\}\, .
\end{eqnarray}
Let us take the expectation with respect to the 
signatures $\sig$, and normalize by the number of users.
Using the all $+1$ assumption, we get (derivative and expectation can be 
interchanged because $H(X|Y)$ has positive bounded derivative,
see below)
\begin{align}
\frac{1}{K}\frac{\de \E H(X|Y)}{\de \sigma^2}
&= \frac{1}{2\sigma^4}\cdot\label{eq:GEXITDefinition}\\
&\cdot\frac{1}{N\la\alpha}\sum_{a=1}^N\E^+_{y,\sig}\left\{|\partial a|
-\Big(\sum_{i\in\partial a}s_{ia}\ex_i\Big)^2\right\}\, ,\nonumber
\end{align}
where $\ex_i = \ex_i(Y,\sig) \equiv \E[X_i|Y,\sig]$. We shall sometimes refer
to the right hand side as to the GEXIT function and
denote it by $\gexit_{N}(\alpha,\sigma^2)$. From the above expressions it is
easy to realize that  $0\le\gexit_{N}(\alpha,\sigma^2)\le 1/2\sigma^4$.
The same inequalities also hold at fixed $\sig$, which 
justifies the exchange of derivative and expectation above.

As in Refs.~\cite{Life,BECArea,GeneralArea}, we introduce furthermore
the BP GEXIT function $\gbp^t(\alpha,\sigma^2)$, with $t$ a non-negative
integer. This is defined by replacing the expectation 
$\sum_{i\in\partial a}s_{ia}\ex_i$ on the right hand side of 
Eq.~(\ref{eq:GEXITDefinition}) by its estimate after $t$ iterations
of BP (in the $N\to\infty$ limit). In terms of the DE
variables
\begin{eqnarray}
\gbp^t(\alpha,\sigma^2) = \frac{1}{2\sigma^4\la\alpha}\E
\left\{k-\left\<\sum_{i=1}^ks_i\xi_i\right\>^2\right\}\, ,
\end{eqnarray}
where $\<\;\cdot\;\>$ denotes an average over $\xi_i\in\{+1,-1\}$ with
distribution
\begin{eqnarray}
\nu(\{\xi_i\}) = \frac{1}{\Xi}\,
e^{-\frac{1}{2\sigma^2}\left(w+\frac{1}{\sqrt{\la}}\sum_{i=1}^k
s_i(1-\xi_i)\right)^2}\;\prod_{i=1}^ke^{v_i^t\xi_i}\, ,
\end{eqnarray}
and the expectation $\E$ is taken with respect to $\{v_i^t\}$
(i.i.d. and distributed as $v^t$ from DE),
$\{s_i\}$ (i.i.d. uniform in $\{+1,-1\}$), $w$ (gaussian with mean zero
and variance $\sigma^2$), and $k$ (Poisson with mean $\la\alpha$).
%
%
\subsection{The proof}

The proof of Theorem \ref{thm:Entropy} makes use of three lemmas,
which we state without demonstration for lack of space.
As in Section \ref{ResultsSection}, $\hs(\alpha,\sigma^2,\la)$
and  $\hi(\alpha,\sigma^2,\la)$ denote, respectively, the $\limsup$
and $\liminf$ of the expected conditional entropy per bit, in the system 
with Poisson signatures.

The first lemma states that, in the low noise limit, the input can be 
reconstructed faithfully from the transmitted message and therefore the
conditional entropy per bit vanishes (recall that we are dealing with discrete
inputs).
\begin{lemma}\label{lemma:ZeroNoise}
For any $\alpha>0$,
$\lim_{\sigma^2\to 0}\lim_{\la\to\infty}\hs(\alpha,\sigma^2,\la)= 0$.
\end{lemma}
The proof is based on a union bound, and a combinatorial calculation.

The second lemma provides upper and lower bounds
on the conditional entropy per user, in terms of BP GEXIT functions.
For the sake of definiteness, we state the lemma for Poisson signatures 
(and denote the corresponding BP GEXIT functions as 
$\gbp^t(\alpha,{\sigma'}^2,\la)$)
although it obviously holds in greater generality \cite{GeneralArea}.
\begin{lemma}\label{lemma:GBP}
For any $\la> 0$, $\sigma_0^2>0$, and non-negative integer $t$
\begin{align}
1-&\int_{\sigma^2}^{\infty}\gbp^t(\alpha,{\sigma'}^2,\la)\;\de{\sigma'}^2
\le\hi(\alpha,\sigma^2,\la)\le\label{eq:GBPBound}\\
&\le\hs(\alpha,\sigma^2,\la) 
\le\hs(\alpha,\sigma_0^2,\la)+\int_{\sigma_0^2}^{\sigma^2}
\gbp^t(\alpha,{\sigma'}^2,\la)\;\de{\sigma'}^2\, .\nonumber
\end{align}
\end{lemma}
This is in fact an easy consequence of the general result that 
GEXIT functions preserve physical degradation \cite{GeneralArea}.

Finally, a Lemma on the large $\la$ limit of DE.
\begin{lemma}\label{lemma:LargeLav}
Define the sequence $\{\lambda_t;\; t\ge 0\}$ by setting $\lambda_0=0$
and
\begin{eqnarray}
\lambda_{t+1} = 
\left\{\sigma^2 +\alpha\left[1-
\Ez \tanh^2(\lambda_t+\sqrt{\lambda_t}\, z)\right]\right\}^{-1}\, ,
\label{lambdaDef}
\end{eqnarray}
for any $t\ge 0$. Let $\{v^t;\; t\ge 0\}$ 
be the solution of DE for the system with 
Poisson signatures (with mean $\la$) and
the same values of $\sigma^2$ and $\alpha$. Then, for any $t\ge 0$, $v^t$
converges in distribution  to a gaussian random variable with
mean $\lambda_t$ and variance $\lambda_t$  as $\la\to\infty$.
\end{lemma}
The proof is based on a repeated application of the central limit
theorem (the argument can be written as an induction over $t$).
The reader is invited to try, for instance, with $t=1,2,\dots$. 

Let us now turn to the  proof of Theorem \ref{thm:Entropy}. We start
by using Lemma \ref{lemma:LargeLav} to compute the large $\la$
limit of the BP GEXIT functions. After a simple application of
central limit theorem, we get 
\begin{eqnarray}
\lim_{\la\to\infty} \gbp^t(\alpha,{\sigma}^2,\la) = \frac{1}{2\sigma^2}
\,\frac{(1-q_t)}{\sigma^2+\alpha(1-q_t)}\, ,\label{eq:AsymptoticGexit}
\end{eqnarray}
where  $q_t \equiv \Ez \tanh^2(\lambda_t+\sqrt{\lambda_t}\, z)$.
We shall denote the expression on the right hand side of 
Eq.~(\ref{eq:AsymptoticGexit}) as $\gbp^t(\alpha,{\sigma}^2)$.

\begin{figure}[t!]
\center{\includegraphics[width=9cm]{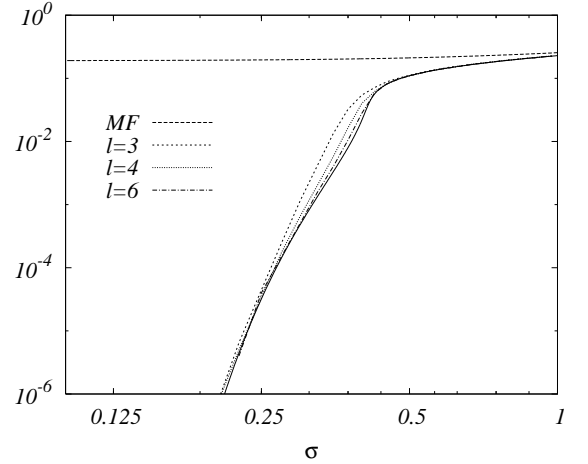}}
\caption{The bit error rate as a function of the noise parameter 
$\sigma$ at $\alpha=1.3$. 
The bold continuous line is Tanaka's result for dense signatures
under symbol MAP detection. MF refer to the same signature scheme under
matched filter detection. The other (dashed) lines correspond to
sparse signatures and BP detection.}
\label{fig:a13}
\end{figure}
Next, we use Lemma \ref{lemma:GBP}. Noticing that  
$0\le \gbp^t(\alpha,{\sigma}^2,\la)\le 1/4\sigma^2$ we
can apply the dominated convergence theorem to take the $\la\to\infty$
limit in Eq.~(\ref{eq:GBPBound}). If we take $\sigma_0^2\to 0$
afterwards and apply Lemma \ref{lemma:ZeroNoise}, we get
\begin{align}
1-\int_{\sigma^2}^{\infty}&\gbp^t(\alpha,{\sigma'}^2)\;\de{\sigma'}^2
\le\hi(\alpha,\sigma^2,\infty)\le\label{eq:BoundLinfty}\\
&\le\hs(\alpha,\sigma^2,\infty) \le \int_{0}^{\sigma^2}
\gbp^t(\alpha,{\sigma'}^2)\;\de{\sigma'}^2\, ,\nonumber
\end{align}
where
$\hi(\alpha,\sigma^2,\infty)\equiv\liminf_{\la\to\infty}
\hi(\alpha,\sigma^2,\la)$, and
$\hs(\alpha,\sigma^2,\infty)\equiv$ $\limsup_{\la\to\infty}
\hs(\alpha,\sigma^2,\la)$.

Simple calculus shows that $\lambda_t$ is strictly positive and increasing in 
$t$  for $t\ge 1$, and $\lambda_t\simeq \sigma^{-2}$ as $\sigma\to 0$.
Furthermore $\lim_{t\to\infty}\lambda_t =\lbp$ is the 
smallest positive fixed point of the recursion (\ref{lambdaDef}),
i.e. the smallest positive solution of  Tanaka's stationarity
equation (\ref{eq:SaddlePoint}). 

From these remarks, it follows that $\gbp^t(\alpha,\sigma^2)$
is integrable over $\sigma\in[0,\infty)$ and strictly decreasing in 
$t\ge 1$. We can therefore take the $t\to\infty$ limit of 
Eq.~(\ref{eq:BoundLinfty}) to get
\begin{align}
1-\int_{\sigma^2}^{\infty}&\gbp(\alpha,{\sigma'}^2)\;\de{\sigma'}^2
\le\hi(\alpha,\sigma^2,\infty)\le\label{eq:BoundLinfty2}\\
&\le\hs(\alpha,\sigma^2,\infty) \le \int_{0}^{\sigma^2}
\gbp(\alpha,{\sigma'}^2)\;\de{\sigma'}^2\, ,\nonumber
\end{align}
where we defined
\begin{eqnarray}
\gbp(\alpha,{\sigma}^2)\equiv \lim_{t\to\infty} 
\gbp^t(\alpha,{\sigma}^2)= \frac{1}{2\sigma^2}\;
\frac{1-\qbp}{\sigma^2+\alpha(1-\qbp)}\nonumber\, ,
\end{eqnarray}
and $\qbp = \Ez\tanh(\lbp+\sqrt{\lbp}z)$. 

We are left with the task of  showing that the first and the last
expressions in Eq.~(\ref{eq:BoundLinfty2}) do indeed coincide and are 
both equal to Tanaka's formula $\hrs(\alpha,\sigma^2)$. 
Recall that, for $\alpha<\alpha_{\rm s}$, the stationarity
equation (\ref{eq:BoundLinfty}) admits a unique solution  depending 
smoothly on $\sigma^2$. Furthermore, we saw above that this coincides 
with the BP fixed point. Using these remarks, we can differentiate 
Eq.~(\ref{hrs}) with respect to $\sigma^2$, to get
\begin{eqnarray}
\frac{\partial \hrs}{\partial\sigma^2}(\alpha,\sigma^2) = 
\gbp(\alpha,{\sigma}^2)\, .
\end{eqnarray}
The proof is completed by applying the fundamental theorem of calculus
to Eq.~(\ref{eq:BoundLinfty2}) and noticing that $\hrs(\alpha,0) = 0$ and
$\hrs(\alpha,\infty) = 1$.\endproof
%
%
\section{Numerical simulations}
\label{NumericalSection}

\begin{figure}[t!]
\center{\includegraphics[width=9cm]{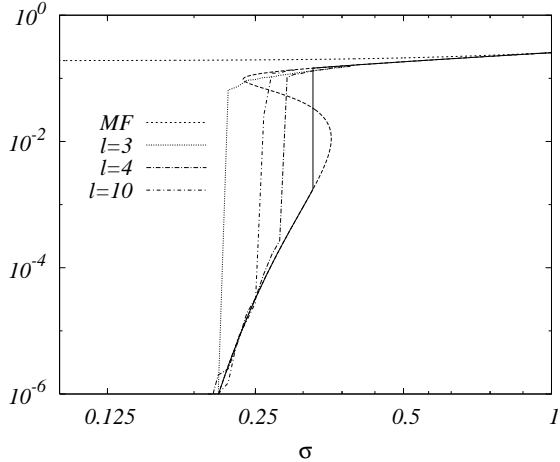}}
\caption{Same as in Fig.~\ref{fig:a13} but for $\alpha=1.9$. The S
shaped dashed curve is the analytical continuation of 
the bit error rate for dense signatures.}
\label{fig:a19}
\end{figure}

One may wonder how quickly is the $\la\to\infty$ limit in Theorems 
\ref{thm:Entropy} and \ref{thm:BP} attained. In Fig.~\ref{fig:a13} we show
the results of numerical simulations using DE, and
regular signatures ($\Omega_l$ concentrated on a single value),
for $\alpha=1.3<\alpha_{\rm s}$.
Already at $l=4$ the bit error rate is extremely close to the
dense limit!

Even more surprising is the behavior for $\alpha>\alpha_{\rm s}$.
In Fig.~\ref{fig:a19} we show the data for $\alpha=1.9$. The BP
error rate at $l=4$ is close to the MAP one with dense signatures.
However it worsens at $l$ grows (and seems to approach the natural 
guess for BP behavior with dense signatures). Sparse signatures
are the crucial ingredient allowing for low complexity detection
and close-to-optimal performances.
%
%

\end{document}